\documentclass{article}
\usepackage{amsfonts}
\usepackage{spconf,amsmath,graphicx,booktabs,mathtools}
\usepackage{CJKutf8}          
\usepackage{url}
\usepackage{microtype}         
\usepackage[hidelinks]{hyperref} 
\usepackage{mathtools}
\usepackage{amsmath}
\usepackage{breqn}
\usepackage{graphicx}
\usepackage{booktabs}
\usepackage{float}
\usepackage{cuted}
\usepackage{capt-of}
\usepackage[table]{xcolor}
\usepackage{siunitx}
\definecolor{chBlue}{HTML}{4E80EE}

\title{No Verifiable Reward for Prosody: Toward Preference-Guided Prosody Learning in TTS}
\name{
  Seungyoun Shin\textsuperscript{1},
  Dongha Ahn\textsuperscript{2},
  Jiwoo Kim\textsuperscript{3,*}\thanks{* Work done during internship at Channel Corporation.},
  Sungwook Jeon\textsuperscript{4,\textdagger}\thanks{\textdagger\; Corresponding author.}
}
\address{
    $^{1}$Channel Corporation, Seoul, South Korea \\
    $^{2}$Kernelspace, Seoul, South Korea \\
    $^{3}$Sungkyunkwan University, Suwon, South Korea \\
    $^{4}$NAVER Cloud, South Korea
}

\begin{document}
%
\maketitle


\begin{abstract}
Recent work reports gains in neural text-to-speech (TTS) with Group Relative Policy Optimization (GRPO). However, in the absence of a verifiable reward for \textit{prosody}, GRPO trained on transcription-oriented signals (CER/NLL) lowers error rates yet collapses prosody into monotone, unnatural speech; adding speaker-similarity further destabilizes training and degrades CER. We address this with an \textit{iterative Direct Preference Optimization (DPO)} scheme that uses only a few hundred human-labeled preference pairs per round to directly optimize prosodic naturalness while regularizing to the current model. On \textbf{KoCC-TTS}, a curated dataset of authentic Korean call center interactions capturing task-oriented dialogues, our method attains the highest human preference (ELO) with competitive CER, outperforming GRPO and strong commercial baselines. These results suggest that when prosody cannot be rewarded automatically, \textit{human preference optimization} offers a practical and data-efficient path to natural and robust TTS. The demo page is available at \href{https://tts.ch.dev}{https://tts.ch.dev}.

\end{abstract}

\begin{keywords}
text-to-speech, prosody, naturalness, preference optimization, verifiable reward
\end{keywords}

\section{Introduction}
\label{sec:intro}

Recent advances in neural text-to-speech (TTS) have achieved near-human intelligibility with autoregressive and diffusion models \cite{shen2018natural,du2024cosyvoice,mehta2024matcha,ye2025llasa}. However, \emph{prosodic control} remains challenging: state-of-the-art systems still struggle to render natural pitch movement and phrasing in conversational settings \cite{shim2025prosody}. At the same time, reinforcement learning has emerged as a promising approach for aligning generated speech with desired attributes, whose effectiveness depends on the design of the reward signal \cite{li2025dmospeech2, atamanenko2025tts1, rafailov2023direct}. 

We argue that the lack of naturalness stems from a gap in reward design. Reliable automatic metrics for prosody remain limited, making it difficult to provide reinforcement signals that align with natural speech patterns.  Optimizing GRPO on CER or NLL improves intelligibility but suppresses prosodic variation, often resulting in near-monotone speech. However incorporating speaker-similarity rewards further introduces instability and degrades performance, inflating CER. In this paper, we contend that the bottleneck lies in the reward formulation rather than in the choice of optimizer. 

\begin{figure}[t]
  \centering
  \includegraphics[width=\columnwidth]{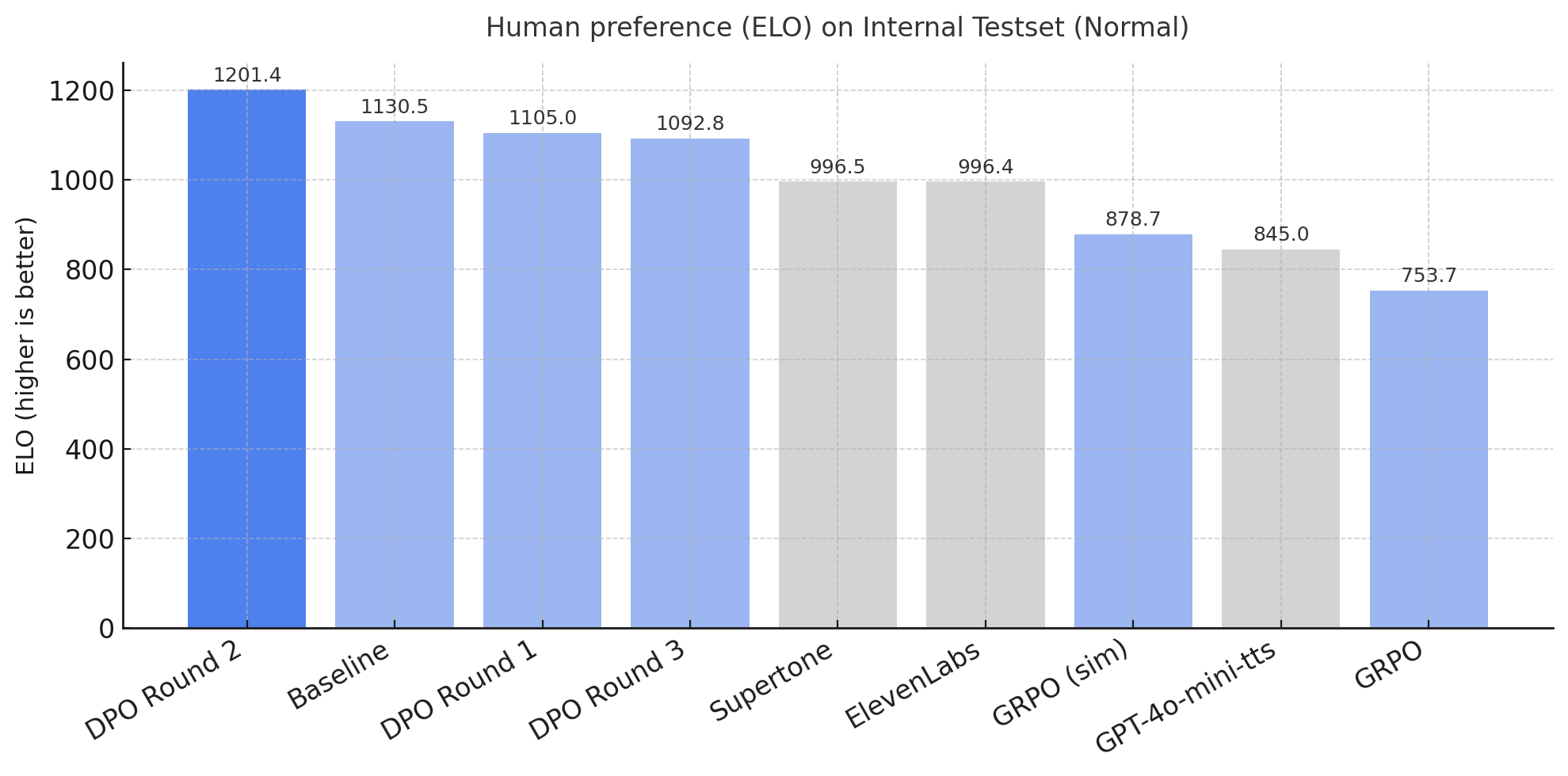}
  \caption{Human preference (ELO) on \texttt{KoCC-TTS}. Iterative DPO (Round~2) ranks highest with GRPO the lowest.}
  \label{fig:elo_score}
\end{figure}

To close this reward gap, we adopt iterative Direct Preference Optimization (DPO) with small human-in-the-loop batches. Across rounds, DPO supplies a directly verifiable signal for prosodic naturalness while regularizing to the current model, yielding both improved human preference and competitive CER (Fig.~\ref{fig:elo_score}). Our contributions are as follows:

\begin{itemize}
  \item We identify a reward-design failure for prosody: CER/NLL-driven GRPO collapses pitch and phrasing, and adding speaker similarity destabilizes training.
  \item We demonstrate that applying \emph{iterative Direct Preference Optimization (DPO)} with $\sim$200 human preference pairs per round restores conversational prosody while keeping CER competitive.
\item We release \texttt{KoCC-TTS}, an evaluation set of frequent Korean call-center phrases for TTS prosody assessment.\footnote{\url{https://huggingface.co/datasets/channelcorp/KoCC-TTS-testset}}
\end{itemize}

\section{Related Work}
\label{sec:related}

\textbf{GRPO for TTS and reward design}
Recent TTS studies that adopt group-relative policy optimization (GRPO) primarily target intelligibility and identity preservation by rewarding ASR-derived errors and speaker similarity, sometimes adding non-intrusive quality predictors.
For instance, F5R-TTS couples WER with speaker-similarity (SIM) under GRPO \cite{sun2025f5rtts}, DMOSpeech2 optimizes a duration policy with SIM+WER via GRPO \cite{li2025dmospeech2}, and the TTS-1 technical report describes a composite GRPO reward that blends WER/SIM/DNSMOS for RL alignment \cite{atamanenko2025tts1}.
While these report lower error rates and stronger speaker faithfulness, they largely omit explicit prosody-sensitive rewards (e.g., pitch movement, phrasing, boundary control).
In practice, we also observed that CER/NLL oriented GRPO can collapse prosodic variation into near-monotone renderings, consistent with reports of punctuation and phrasing related failures in state-of-the-art systems \cite{shim2025prosody}.

\textbf{Preference based objectives for prosody}
Direct Preference Optimization (DPO) offers a complementary route by optimizing pairwise human (or proxy) preferences without training a separate reward model \cite{rafailov2023direct}.
In TTS, Emo-DPO applies DPO to better capture subtle emotional/prosodic nuances with an LLM-based TTS backbone, improving both prosody similarity and perceived naturalness \cite{gao2024emodpo}.
Beyond single shot DPO, iterative preference optimization for speech generation has also been explored. SpeechAlign constructs codec-token preference pairs and refines a speech LM in multiple rounds, demonstrating iterative self-improvement \cite{zhang2024speechalign}.
Concurrently, differentiable or multi-dimensional preference objectives have been proposed to move past coarse ASR metrics. DiffRO directly optimizes differentiable rewards over codec tokens \cite{gao2025diffro}, and MPO considers multi-criteria screening of preference pairs in speech synthesis \cite{zhou2025mpo}.
These studies collectively suggest that \emph{preference based post training} is a promising way to recover communicative prosody without sacrificing robustness to transcription errors.

\section{Methodology}
\label{sec:method}

\subsection{Training Data}
\label{subsec:training-data}
We employ approximately 36k hours of publicly available Korean \texttt{(text, audio)} pairs from AIHUB.\footnote{\url{https://aihub.or.kr/}} 
In addition, we curate 18 hours of proprietary single-speaker data (female voice) consisting of manager–customer dialogues. 
Only the manager channel is retained to ensure consistent speaker characteristics. 
Speech-active regions are extracted using \texttt{pyannote.audio}~\cite{Bredin23} (v3.0) and transcribed with Whisper-large-v3, producing segmented training pairs of the form \texttt{[(audio\_1, text\_1), (audio\_2, text\_2), ...]}.

\subsection{Base Model}
\label{subsec:base-model}

We adopt an architecture based on Llasa, which uses a Transformer (initialized from LLaMA) to generate discrete speech tokens decoded into waveforms via XCodec2\cite{ye2025llasa}. Starting from the Llasa-1B checkpoint,\footnote{\url{https://huggingface.co/HKUSTAudio/Llasa-1B}} we perform continual training on a 36k hours Korean corpus to instill language-specific competence. Then we fine-tune on an 18-hour proprietary single-speaker dataset to adapt prosody toward a natural conversational style. We refer to this model as \texttt{channel-base}.

\subsection{Reinforcement Learning with GRPO}
\label{subsec:grpo}

We employ Group Relative Policy Optimization (GRPO), a PPO-style variant that
optimizes grouped samples without an explicit value network~\cite{shao2024deepseekmath,guo2025deepseek}.

\subsubsection{Base reward}

\vspace{0.5em}
\noindent\textbf{Notation}
We denote by $c \ge 0$ denote the character error rate (CER) computed by ASR on the synthesized audio
, which may exceed 1 due to insertions, and by $\ell \ge 0$ denote the average negative log-likelihood (NLL) per generated token.
We further introduce temperature parameters $\tau_c,\tau_\ell>0$ and reward weights $\lambda_c,\lambda_\ell>0$, normalized without loss of generality such that $\lambda_c+\lambda_\ell=1$.

\vspace{0.35em}
\noindent\textbf{Utilities}
We map each metric to $(0,1]$:
\begin{equation}
U_{c} = 1 - \mathrm{tanh}(\tau_c\, c),
\qquad
U_{\ell} = \exp\!\left(-\frac{\ell}{\tau_\ell}\right).
\label{eq:utilities}
\end{equation}
Since $U_c,U_\ell \in (0,1]$, the reward below also lies in $(0,1]$.

\vspace{0.35em}
\noindent\textbf{Reward}
\begin{equation}
R =
\frac{\lambda_c+\lambda_\ell}{\lambda_c/U_{c} + \lambda_\ell/U_{\ell}}
\;\in\; (0,1].
\label{eq:hm-2term}
\end{equation}
The harmonic mean penalizes small components, creating strong pressure against high error while still rewarding acoustic likelihood.

\vspace{0.35em}
\noindent\textbf{Settings}
Empirically, we set $(\lambda_c,\lambda_\ell)=(0.6,\,0.4)$; $(\tau_c,\tau_\ell)$ are tuned on a held-out development set.

\subsubsection{Speaker similarity extension}

\vspace{0.5em}
\noindent\textbf{Utility}
To encourage target-speaker faithfulness, we introduce a speaker-similarity utility.
Let $s\in[-1,1]$ denote the cosine similarity between speaker embeddings. 
We map $s$ into $(0,1]$ via an elementwise clamp:
\begin{equation}
U_{s} \;=\; \min\!\big(\max\!\big((s+1)/2,\,0\big),\,1\big).
\label{eq:sim-utility}
\end{equation}

\vspace{0.35em}
\noindent\textbf{Reward}
With positive weights $\lambda_c,\lambda_\ell,\lambda_s$ (normalized such that $\lambda_c+\lambda_\ell+\lambda_s=1$), 
the training reward is defined as
\begin{equation}
R \;=\;
\frac{\lambda_c+\lambda_\ell+\lambda_s}
{\lambda_c/U_{c} \;+\; \lambda_\ell/U_{\ell} \;+\; \lambda_s/U_{s}}
\;\in\; (0,1].
\label{eq:hm-3term}
\end{equation}

\vspace{0.35em}
\noindent\textbf{Settings}
In our experiments, we use $(\lambda_c,\lambda_\ell,\lambda_s)=(0.5,\,0.3,\,0.2)$;
$(\tau_c,\tau_\ell)$ follow the two-term setup.

\subsection{Iterative Direct Preference Optimization (DPO) for Prosody}
\label{subsec:dpo}

To restore prosodic variation while preserving transcription robustness, we perform \emph{round-based} preference learning with Direct Preference Optimization (DPO)~\cite{rafailov2023direct}. 
At round $r\in\{1,2,3\}$, the policy is initialized from the previous checkpoint $\pi_{\theta_{r-1}}$, which also serves as the moving reference $\pi_{\mathrm{ref}}=\pi_{\theta_{r-1}}$. 
We generate candidates with $\pi_{\theta_{r-1}}$, collect 200 human preference pairs $\{(x,y^+,y^-)\}$, and update the policy by optimizing the DPO objective to obtain $\pi_{\theta_r}$. 
This procedure yields \texttt{channel-base-dpo-v1}, \texttt{channel-base-dpo-v2}, and \texttt{channel-base-dpo-v3} for $r=1,2,3$, respectively. 
Preference data are not reused across rounds.

\vspace{0.5em}
\noindent\textbf{Objective}
Following Rafailov et al.~\cite{rafailov2023direct}, the log-likelihood gaps are defined as
\begin{align}
\Delta \ell_\theta(x,y^+,y^-) 
&:= \log \pi_\theta(y^+\!\mid x)
   - \log \pi_\theta(y^-\!\mid x), \\
\Delta \ell_{\mathrm{ref}}(x,y^+,y^-) 
&:= \log \pi_{\mathrm{ref}}(y^+\!\mid x)
   - \log \pi_{\mathrm{ref}}(y^-\!\mid x).
\end{align}

The DPO loss is then given by~\cite{rafailov2023direct}
\begin{equation}
\resizebox{\columnwidth}{!}{$
\mathcal{L}_{\mathrm{DPO}}(\theta)
= -\mathbb{E}_{(x,y^+,y^-)} \left[
\log \sigma\!\left( \beta\!\left[
\Delta \ell_\theta(x,y^+,y^-)-\Delta \ell_{\mathrm{ref}}(x,y^+,y^-)
\right]\right)\right]
$}
\end{equation}
\label{eq:dpo}

where $\sigma(\cdot)$ is the logistic function and $\beta>0$ controls preference sharpness.
This objective increases the likelihood ratio of preferred over dispreferred outputs while implicitly regularizing toward the round-specific reference, which we target at prosodic naturalness.

\section{Experiments}
\label{sec:exp}

\begin{table}[t]
\centering
\small
\setlength{\tabcolsep}{6pt}
\renewcommand{\arraystretch}{1.12}
\begin{tabular}{l S[table-format=2.2] c}
\toprule
\textbf{Model} & {\textbf{CER} $\downarrow$ (\%)} & {\textbf{ELO}} \\
\midrule
ElevenLabs (Multilingual v2)\footnotemark[1] & 4.74 & 955.1 \\
Supertone\footnotemark[2]                    & 2.98 & 1046.9 \\
GPT\mbox{-}4o\mbox{-}mini\mbox{-}tts (sage)  & 2.91 & 848.9 \\
Llasa-8B                                     & 3.24 & -- \\
Llasa-3B                                     & 3.47 & -- \\
Llasa-1B                                     & 10.45 & -- \\
\addlinespace[2pt]
\multicolumn{3}{l}{\textbf{Ours}} \\
\cmidrule(l){1-3}
\texttt{channel-base}          & 2.90 & 1150.1 \\
GRPO (clean)                 & \textbf{2.20} & 753.7 \\
GRPO\mbox{-}sim extension    & 42.63 & 878.7 \\
\texttt{channel-base-dpo-v1}                 & 5.80 & 1096.5 \\
\texttt{channel-base-dpo-v2}              & 3.60 & \cellcolor{chBlue!22}1190.1 \\
\texttt{channel-base-dpo-v3}             & 3.30 & 1064.2 \\
\bottomrule
\end{tabular}
\caption{Results on \texttt{KoCC-TTS}. CER (\%, lower is better) and ELO-based human preference (higher is better). Rows under \textbf{Ours} are internal models; the shaded entry marks the best DPO round (R2).}

\label{tab:cer_elo}
\end{table}

\subsection{KoCC-TTS}
\label{subsec:setup}
We constructed a new dataset, \texttt{KoCC-TTS}(Korean Call-Center TTS), consisting of 50 high quality human-curated samples drawn from real manager–user conversations. This dataset provides challenging, domain-specific utterances, serving as a reliable testbed to evaluate transcription robustness as well as conversational prosody in Korean task-oriented speech synthesis.

\subsection{Setup}
We evaluate 12 systems on the KoCC-TTS dataset, including 3 production-grade external services, 3 open-source models, and 6 internal variants. We intentionally exclude open-source TTS baselines from the main comparison. A preliminary screening indicated that most off-the-shelf OSS voices exhibited inadequate prosodic fluency in Korean, and their inclusion would likely result in floor effects rather than provide a meaningful basis for comparison. 
For external systems, we adopt vendors' strongest Korean voices and default synthesis settings unless otherwise noted: ElevenLabs Multilingual v2 ("Anna")\footnote{\url{https://elevenlabs.io/blog/eleven-multilingual-v2}}, Supertone ("\texttt{sona\_speech\_1}")\footnote{\url{https://docs.supertoneapi.com/ko/user-guide/quickstart}}, and GPT-4o-mini-tts ("sage"). To ensure fairness, all systems synthesize from the same prompts with identical text normalization rules. Speaking rate and punctuation handling are held fixed, and outputs are evaluated at vendors’ native sampling configurations.

We report (i) character error rate (CER) computed from Whisper-large-v3 transcriptions and (ii) human preference aggregated into ELO scores. For human evaluation, we adopt blind A/B pairwise comparison following Chatbot Arena-style evaluation~\cite{chiang2024chatbot}. We collected 596 votes from 27 participants whose ages ranged from 20 to 60. In each trial, raters listened to two anonymized audio samples and selected win, loss, or tie based on which utterance sounded more natural in terms of pitch and prosodic flow. Votes were aggregated via ELO-style ranking. Aggregate results are summarized in Table~\ref{tab:cer_elo}, with ELO rankings visualized in Figure~\ref{fig:elo_score}.

\footnotetext[1]{\url{https://elevenlabs.io/blog/eleven-multilingual-v2}}
\footnotetext[2]{\url{https://www.supertone.ai/ko}}

\subsection{GRPO and Prosodic Collapse}
\label{subsec:grpo-prosody}

Applying the reward function in Eq.~\ref{eq:hm-2term}, GRPO consistently reduces CER to the lowest level among all variants. All GRPO models were trained on 1.6M text prompts.
However, as illustrated in Fig.~\ref{fig:prosody}, the ${\log}$F0 distribution of GRPO-trained models shows reduced pitch variability compared to the baseline, indicating a collapse toward monotonic prosody. 
Although such optimization improves transcription robustness, it results in speech that listeners perceive as unnatural, which explains the lower ELO scores relative to CER gains.

\subsection{Speaker-Similarity Extension and Training Instability}
\label{subsec:grpo-sim}

To address monotonicity, we introduced an additional speaker-similarity term in the reward (Eq.~\ref{eq:hm-3term}). 
While this modification increased similarity scores, it also destabilized training: CER degraded substantially, and we observed degenerate behaviors where the model generated excessively long outputs without producing an end-of-sequence token. 
Although the text was realized, utterances frequently failed to terminate, suggesting that the RL objective was partially ``hacked.'' 
These results indicate that incorporating speaker-similarity rewards into GRPO introduces optimization challenges and reduces training stability, making it unsuitable as a direct solution for prosodic control.

\subsection{Iterative DPO: Small Preference Sets Recover Prosody \& CER}
\label{subsec:iter-dpo}

We next apply round-based Direct Preference Optimization (DPO) with 200 human-labeled pairs per round, using a moving reference
(\(\pi_{\mathrm{ref}}=\pi_{\theta_{r-1}}\)) and no replay from earlier rounds.
Each round regenerates candidates, collects fresh A/B preferences, and optimizes Eq.~\eqref{eq:dpo}.

\vspace{0.5em}
\noindent\textbf{Outcomes}
As summarized in Table~\ref{tab:cer_elo}, starting from \texttt{channel-base} (CER = 2.90\%, ELO = 1150.1),
GRPO attains the lowest CER (2.20\%) but the lowest preference (ELO = 753.7) due to monotone prosody.
Iterative DPO reverses this trade-off:
\begin{itemize}
  \item \textbf{Round~1:} ELO rises to 1096.5 while CER increases to 5.80\% as the model explores more varied prosody.
  \item \textbf{Round~2:} ELO peaks at \textbf{1190.1} and CER improves to 3.60\%, outperforming external systems in preference and approaching baseline CER.
  \item \textbf{Round~3:} CER further improves to 3.30\% with ELO = 1064.2, retaining a clear prosodic advantage over GRPO.
\end{itemize}

\vspace{0.35em}
\noindent\textbf{Why does Round~2 peak?}
We hypothesize that early rounds benefit from larger reward gaps between chosen and rejected samples, providing more informative gradients for preference learning. As iterations proceed, the policy-reference gap narrows and the marginal informativeness of new preference pairs diminishes, leading to saturation. This pattern is consistent with diminishing returns observed in iterative preference optimization~\cite{pang2024iterative}.

\begin{figure}[t]
\centering
\includegraphics[width=8.5cm]{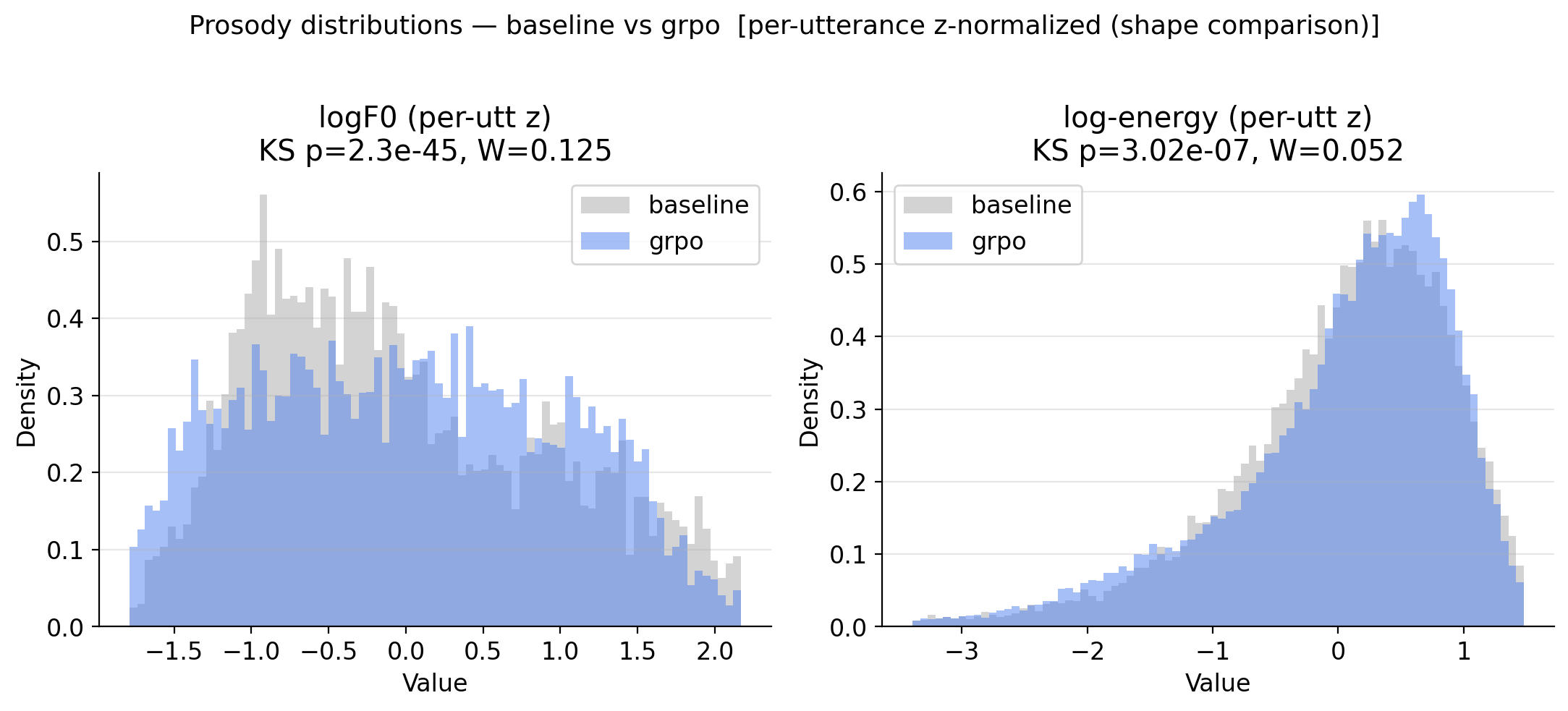}
\caption{Pitch contour (logF0) distribution before and after GRPO. The baseline corresponds to \texttt{channel-base}.}
\label{fig:prosody}
\end{figure}

\vspace{0.35em}
\noindent\textbf{Takeaways}
With only 200 pairs per round, iterative DPO (i) restores prosodic variation favored by listeners, as reflected in higher ELO scores, and (ii) reduces CER after the initial exploration phase.
As shown in Fig.~\ref{fig:prosody} (increased $F_0$ variability compared with GRPO), these results indicate that preference learning complements GRPO by mitigating prosodic collapse while maintaining competitive transcription robustness.

\section{Conclusion}
\label{sec:conclusion}

Without a verifiable automatic reward for prosody, GRPO trained on transcription-centric signals (CER/Whisper-NLL) predictably optimizes what is measured—intelligibility—while collapsing what is not—prosodic variation—into near-monotone speech. Extending the reward with speaker-similarity injects noisy, non-prosodic supervision that destabilizes optimization (e.g., EOS failures) and inflates CER, indicating that the core limitation lies in the reward design, not the optimizer.

We close this reward gap with iterative DPO, replacing unverifiable proxies with directly verifiable human preferences. With only 200 preference pairs per round, DPO consistently restores prosodic diversity favored by listeners (highest ELO) while keeping CER competitive, serving as a data-efficient complement to GRPO. We also release \textbf{KoCC-TTS} for robustness and conversational prosody evaluation. Our takeaway is simple: \emph{when prosody cannot be reliably rewarded automatically, human-in-the-loop preference optimization is the practical path to natural and robust TTS}.

\section{Acknowledgements}
We thank the Channel Corporation for providing
GPU resources to run the experiments, and AI
team for providing helpful feedback.  

\bibliographystyle{IEEEbib}
\bibliography{refs}

\end{document}